\documentstyle[12pt]{article}
\iftrue \textheight44\baselineskip 
\else\textheight22.42cm \fi
\begin{document}
\title{Dynamics of Homogenous Magnetizations in Strong Transverse Driving
Fields}
\author{Thomas Tr\"axler, Wolfram Just, and Herwig Sauermann\\
Technische Hochschule Darmstadt\\Institut f\"ur Festk\"orperphysik\\
Hochschulstra\ss e 8\\ 64289 Darmstadt\\ Germany\\
Tel.: + 49 6151 165362\\
FAX + 49 6151 164165\\
e--mail: wolfram@arnold.fkp.physik.th-darmstadt.de}
\maketitle
\begin{abstract}
Spatially homogeneous solutions of the Landau--Lifshitz--Gilbert
equation are
analysed. The conservative as well as the dissipative case is considered
explicitly. For the linearly polarized driven
Hamiltonian system we apply canonical perturbation
theory to uncover the main resonances as well as
the global phase space structure.
In the case of circularly polarized driven dissipative motion we
present the complete bifurcation diagram including bifurcations
up to codimension three.
\end{abstract}
\newpage
\section{Introduction}
The investigation of strongly driven ferromagnetic systems is an interesting
field of research especially from the point of view of nonlinear dynamics.
The fundamental equation of motion, which governs the dynamics of a
macroscopic magnetization has been proposed already
by Landau and Lifshitz \cite{lali}
\begin{equation}\label{diss}
\dot{\bf S}=-{\bf S}\times \left({\bf H}_{eff}({\bf S},t)+ \Gamma {\bf S}
\times {\bf H}_{eff}({\bf S},t)\right)\quad.
\end{equation}
It describes the motion of a macroscopic magnetization ${\bf S}$
under the influence of an effective local magnetic
field ${\bf H}_{eff}({\bf S},t)$.
The second term in this equation, frequently called the Gilbert damping,
represents the dominant contribution among several dissipative
terms, which arise in microscopic derivations of this equation
(e.g. ref.\cite{gara,plef}). It is a special feature of eq.(\ref{diss})
that it preserves the modulus of the magnetization.
In order to give eq.(\ref{diss}) a
definite meaning one has to specify the effective field.
Usually contributions from external fields, dipolar and exchange
interactions, anisotropy etc. are taken into account.
They turn eq.(\ref{diss}) in general into a complicated partial
differential equation which may even be nonlocal in space.
Hence it is impossible to discuss the dynamical behaviour in general.
We will focus in our article on the analysis of spatially homogeneous states.
Even this seemingly simple problem turns out to be highly nontrivial.

To begin with the effective field has to be specified. We take a static
external field, a time dependent transversal driving field and an uniaxial
anisotropy parallel to the static field
into account while contributions from the exchange interaction
obviously vanish and the dipolar interaction may be thought to be contained
in the anisotropy term \cite{Akhi}.
Under these
conditions eq.(\ref{diss}) becomes a two dimensional system of
nonautonomous ordinary differential equations, which is Hamiltonian or
dissipative depending on the value of the damping constant. We will
treat both cases in our paper separately.

Discussions of low dimensional Hamiltonian spin systems can be found
in the literature. A treatment of undriven systems may be found in
\cite{Tho1,Tho2}. Furthermore
Frahm and Mikeska have shown \cite{frmi} that the equations of motion
are integrable, if a circularly polarized driving field is applied to
a single spin.
For linearly polarized driving fields their numerical simulations
indicate that the system is not integrable. In contrast to their
approach we will analyse the Hamiltonian dynamics using canonical
perturbation theory and compare the results with numerical simulations.

On the other hand a systematic analysis of driven dissipative spin systems
does not seem to be available in the literature. We will study the
damped spin system under the influence of a rotating driving field.
In that case the system becomes autonomous using a transformation into
a rotating coordinate system. Although such a system cannot become
chaotic we will show by performing a careful bifurcation analysis that highly
nontrivial phenomena occur.

\section{Hamiltonian Dynamics under Linearly
Polarized Pumping\label{hamlin}}
Undamped spin systems provide a nontrivial example of simple Hamiltonian
systems \cite{frmi}. It is well known that an axisymmetric
system is integrable if a rotating driving field is applied. Hence we
focus in this section on systems which are driven by linearly polarized
fields. They give
rise to very complicated dynamics. To be definite the effective field
is chosen as
\begin{equation}\label{mya}
  {\bf H}_{eff}({\bf S},t)=\left(H_z-a S_z \right)
  {\bf e}_z + h_\bot \cos\left(\omega t\right) {\bf
  e}_x\quad .
\end{equation}
Spin systems of this type have been discussed already
in the literature. But
in contrast to former approaches \cite{frmi,Haak} a static field is
incorporated in our treatment which breaks the plane symmetry $S_z\to -S_z$.
For that reason the methods employed in these references cannot be applied to
our equations of motion.

We will use canonical perturbation theory for weak driving fields to
uncover the phase space structure of our system. The calculations are then
confirmed by numerical simulations.

\subsection{Canonical Perturbation Theory}
The dynamical system (\ref{diss}) subjected to the effective field
(\ref{mya}) is described by the Hamiltonian
\begin{eqnarray}\label{4.5}
{\cal H}&=& {\cal H}_0+ \varepsilon \;{\cal H}_1 \nonumber\\
&=& S_z  \left(H_z - \frac{aS_z}{2} \right) + \varepsilon \;h_{\bot}\:
\sqrt{1-
    S_z^2 } \; \cos (\varphi)\: \sin (\omega t)\quad.
\end{eqnarray}
Here canonical coordinates $S_x + i S_y = \sqrt{1-S_z^2}\exp(i\varphi)$
and the formal expansion
parameter $\epsilon$ have been introduced for convenience.
The unperturbed i.e. the undriven system, $\varepsilon=0$,
is obviously integrable and admits the solution
\begin{eqnarray}
S_z^{(0)}(t) & = & S_z^{(0)}(0)\label{myca} \\
\varphi^{(0)}(t) & = & \varphi^{(0)}(0)+ \Omega(S_z^{(0)}(0)) t
\label{mycb}\\
\Omega(S_z) &:=& \left(H_z - a \;S_z\right) \label{mycc}
\end{eqnarray}
corresponding to the precession of a single spin.

It is the scope of canonical perturbation theory to construct a sequence of
transformations which map the original Hamiltonian approximately to an
integrable one. Formally this is achieved by eliminating fast variables
(in our case $\varphi$ and $t$) up to a certain order
$\varepsilon^m$ in the Hamiltonain .
In each order of this approximation finitely many small denominators will
appear, so that on a finite collection of surfaces the new Hamiltonian is
not well defined. Outside a small neighbourhood of these surfaces
the transformed
system is integrable and may serve as a good approximation to the
original one. The small denominators signal the resonances in the
system and have to be treated separately. They prevent the expansion from
being convergent and even from being valid asymptotically.

We perform the perturbation theory by using a modified version of the von
Zeipel method.
The perturbation expansion is formalized with the help of
the Lie technique\footnote{A good introduction to the
field of Hamiltonian perturbation theory is given in
\cite{LiLi} chapter 2 and \cite{Arno} chapter 5.
For a detailed introduction to the Lie
transformation methods see e.g. \cite{LiLi} section 2.5.}. We are looking
for a canonical transformation $S_z,\varphi\to \bar{S_z},\bar{\varphi}$
which transforms the Hamiltonian (\ref{4.5}) into an integrable one.
Obviously the transformation depends on the expansion parameter
$\varepsilon$.
If $w(J,\varphi,\varepsilon)$ denotes the generator of the infinitesimal
transformation it obeys
\begin{equation}\label{defw}
\frac{d \bar{x}}{d \varepsilon}= \left[ \bar{x},w\right]\quad.
\end{equation}
where $x$ stands for $S_z$ and $\varphi$, respectively and $[.,.]$ denotes
the Poisson bracket. The evolution operator $\bar{x}(\varepsilon)={\bf T} x$
of this formal Hamiltonian system with ''{\it time}'' $\varepsilon$
defines the desired transformation. With the help of the Lie operator
${\bf L}=[w,\quad]$ eq.(\ref{defw}) can be cast into the form
\begin{equation}\label{defT}
\frac{d{\bf T}}{d \varepsilon}=-{\bf TL}\quad .
\end{equation}
This algebraic version of the transformation is well suited for perturbation
expansions. Using the formal power series
\begin{eqnarray}
{\bf L}&=&\sum_{n=0}^\infty \varepsilon^n {\bf L}_{n+1} \label{myba}\\
{\bf T}&=&\sum_{n=0}^\infty \varepsilon^n {\bf T}_{n}\label{mybb}\quad .
\end{eqnarray}
eq.(\ref{defT}) can be solved for ${\bf T}$
\begin{eqnarray}
{\bf T}_n=-\frac{1}{n}\sum_{m=0}^{n-1}{\bf T}_m {\bf L}_{n-m}\label{myia}\\
{\bf T}_n^{-1}=-\frac{1}{n}\sum_{m=0}^{n-1}{\bf L}_{n-m}{\bf T}_m^{-1}
\quad .\label{myib}
\end{eqnarray}
Of course the generator $w$ has to be specified in order
to give these
expressions an explicit meaning. This is accomplished by considering
at first the
transformed Hamiltonian
\begin{equation}
\bar{{\cal H}}(\bar{x}(x,\varepsilon))={\cal H}(x)
\end{equation}
which in terms of the evolution operator (\ref{mybb}) is given by
$\bar{{\cal H}}={\bf T}^{-1}{\cal H}$. From the formal series
expansions
\begin{eqnarray}
{\cal H} &=& \sum_{n=0}^{\infty} \varepsilon^n  {\cal H}_{n} \\
\bar{{\cal H}} &=&\sum_{n=0}^\infty \varepsilon^n \bar{{\cal H}}_{n}
\end{eqnarray}
one obtains the recurrence relations
\begin{equation}\label{myd}
\frac{\partial w_n}{\partial t}+ [w_n,{\cal H}_0]=n
\left(\bar{{\cal H}}_n- {\cal H}_n\right) -
\sum_{m=1}^{n-1} \left({\bf L}_{n-m} \bar{{\cal H}}_m +
m{\bf T}^{-1}_{n-m}{\cal H}_m \right)\quad .
\end{equation}
Owing to the fact, that the unperturbed motion can be integrated
explicitly (cf. eqs.(\ref{myca},\ref{mycb})) these equations
can be solved successively once
$\bar{{\cal H}}_m$ is fixed in an appropriate way. It is
this freedom which is used in the sequel to
construct the perturbation series.
\subsubsection{The First Order}
Let us start with the simple first order perturbation expansion.
Eq.(\ref{myd}) reads explicitly for $n=1$
\begin{eqnarray}\label{ersto}
   \frac{\partial w_1}{\partial t}+ \left[ w_1,{\cal H}_0 \right]&=& \bar{{\cal
       H}}_1 -{\cal H}_1\nonumber\\
   &=& \bar{{\cal H}}_1 - h_\bot\; \sqrt{1-S_z^2}\: \cos (\varphi) \;\sin
       (\omega t)\qquad .
\end{eqnarray}
In order to render the new system integrable one
requires that the new Hamiltonian depends
solely on the action variable. Observing this condition
$\bar{{\cal H}_1}$ is chosen in such a way that
the solution of eq.(\ref{ersto}) possesses no secular contributions.
This is achieved by choosing $\bar{{\cal H}}_1=0 $. We find for
the generator of the transformation
\begin{equation}\label{mys}
   w_1=h_\bot \sqrt{1-S_z^2} \left(\frac{\cos \left(\varphi+ \omega t
          \right)}{2 \left(\Omega + \omega \right)} -
	  \frac{\cos \left(\varphi-\omega t \right)}{2 \left(\Omega -
	  \omega \right)}\right)
\end{equation}
and hence for the transformation itself
\begin{equation}
T f =1 -\varepsilon\; \left[ w_1, f \right] \quad.
\end{equation}
It has already been mentioned that the perturbation expansion leads to
small denominators indicating resonances. They
appear at $\Omega(S_z)=\pm \omega$. Higher
harmonics do not occur in the present case because of the special
nature of the coupling
of the driving field.
\subsubsection{Removal of First Order Resonances}
The local structure of the phase space near
the resonances cannot be described by the approach of the previous paragraph.
However we can uncover the local phase space structure near
some special resonance by
using a different choice of the first order Hamiltonian
$\bar{{\cal H}}_1$. To be
definite let us consider the resonance at $\Omega =-\omega$.
In order to motivate the new choice we transform into a rotating
frame\footnote{It should be
mentioned that the introduction of the rotating frame is by no
means necessary. However the application of these coordinates
is physically more appealing.}
using the new variables
$\hat{S}_z=S_z$ and $\hat{\varphi}=\varphi+\omega t$.
Then the Hamiltonian (\ref{4.5}) reads
\begin{equation}
  \hat{{\cal H}}=\hat{S}_z \left(H_z-\frac{a \hat{S}_z}{2} + \omega
  \right) +\varepsilon\; h_\bot \sqrt{1- \hat{S}_z^2} \left(\sin \left( \hat{
  \varphi} \right) + \sin \left(\hat{\varphi}-2 \;\omega t \right)
  \right)
  \qquad .
\end{equation}
Now
the new angle $\hat{\varphi}$ is a slow variable near the resonance
under consideration. Thus we do not try to remove it from the
Hamiltonian. We are now looking for a solution of
eq.(\ref{myd}) which in the present case reads
\begin{equation}\label{mye}
   \frac{\partial \hat{w}_1}{\partial t}+ \left[ w_1,{\cal H}_0 \right]=
   \bar{\hat{{\cal H}}}_1 - h_\bot\; \sqrt{1-\hat{S}_z^2}\:
   \left(\sin \left( \hat{
    \varphi} \right) + \sin \left(\hat{\varphi}-2 \;\omega t \right)
    \right)
\qquad .
\end{equation}
Choosing
\begin{equation}\label{myr}
\bar{\hat{\cal H}}_1= h_\bot \sqrt{1-\hat{S}_z^2}\;\sin
\hat{\varphi}
\end{equation}
one ensures that the solution
\begin{equation}\label{myx}
  \hat{w}_1=h_\bot \sqrt{1-\hat{S}_z^2}\frac{\cos \left(\hat{\varphi} - 2 \;
     \omega t\right)}{2 \left(H_z- a\hat{S}_z - \omega \right)}
\end{equation}
is regular near the resonance. In addition the new Hamiltonian reads
\begin{equation}\label{fig1}
\bar{\hat{\cal H}}=\hat{S}_z\left(H_z-\frac{a \hat{S}_z}{2} + \omega
\right) +
   \varepsilon \;h_\bot \sqrt{1-\hat{S}_z^2} \sin \hat{\varphi} +
{\cal O} \left(
   \varepsilon^2\right)\label{hamdach}
\end{equation}
is independent of time up to the first order in the
expansion parameter and represents an integrable system. The phase space
structure of this system is sketched in Fig.1\marginpar{$<$ Fig.1}. The system
has a hyperbolic fixed point at $\hat{\varphi}= - \frac{\pi}{2}$ and
an  elliptic fixed point at $\hat{\varphi}=  \frac{\pi}{2}$. In order to
describe the motion near the resonance $\hat{S}_z=(H_z-\omega)/a$,
we expand the Hamiltonian in powers of small deviations $\Delta \hat{S}_z$
from the resonance. In lowest order one obtains the well known
Hamiltonian of a pendulum
\begin{equation}
\bar{\hat{{\cal H}}}=\frac{H_z+\omega}{a}-\frac{a}{2} \left(\Delta
\hat{S}_z
    \right)^2- \varepsilon \;h_\bot \sqrt{ 1 - \left(  \frac{ H_z -
    \omega }{a}\right)^2} \sin \hat{\varphi} \label{pendelgl}\qquad .
\end{equation}
This expression describes the generic case of near resonance motion and is
studied in the literature (e.g. ref.\cite{chir}) very well.
\subsubsection{Second Order Islands}
Taking the higher order contributions of the Hamiltonian
(\ref{pendelgl}) into account leads to perturbation terms of
higher order in $\varepsilon$ and
$\Delta \hat{S}_z$. Especially a time dependence of frequency
$2 \omega$ will enter the Hamiltonian.
We note that the internal frequency of system (\ref{pendelgl}) is of the
order ${\cal O} \left(\sqrt{\varepsilon} \right)$, whereas the external
frequency is of order ${\cal O} \left(1 \right)$. Hence the full system will
develop resonances of order $1/p$, where the integer $p$ is of the order of
magnitude $\varepsilon^{-1/2}$. Now these second order
resonances themselves result in $p$ hyperbolic fixed points and the
heteroclinic tangles belonging to them.
The amplitudes of these resonances is of the order
${\cal O}  \left( 1/\left( \varepsilon^{-1/2}\right)!  \right)$.
For a much more detailed analysis of this case we refer the reader
to the literature (e.g. \cite{LiLi} section 2.4b).
\subsubsection{The Second Order}
The previous paragraphs have dealt with the analysis of
the first order resonances extensively. If we proceed
to higher order additional resonances will occur.
Following the lines of paragraph 2.1.1 we obtain in
second order from eq.(\ref{myd})
\begin{eqnarray}\label{myg}
   \lefteqn{\frac{\partial w_2}{\partial t} + \left[ w_2, {\cal H}_0 \right] =
     2\; \bar{{\cal H}}_2 -\left[w_1,{\cal H}_1\right]}\nonumber\\
    &=&2 \;\bar{{\cal H}}_2-h_\bot^2 a \frac{\left(\cos\left(2 \varphi \right)
       +\cos \left(2 \omega t \right) -1 \right) \left( 1-S_z^2\right)
       \left(\Omega^2  +\omega^2 \right) } {\left( 2
        \left(\Omega^2 -\omega^2\right)\right)^2}\nonumber\\
    &&+h_\bot^2 \frac{S_z\left(1-\cos\left(2 \omega t \right) \right)
       \Omega } {\left( 2 \left(\Omega^2 -
       \omega^2 \right)\right)}\nonumber\\
    &&-h_\bot^2 a \left(1-S_z^2\right)\left(\frac{\cos\left(2 \varphi +2
        \omega t\right) } { 8 \left( \Omega + \omega \right)^2 }
	+\frac{ \cos \left(2 \varphi -2 \omega t \right) } { 8 \left(
         \Omega - \omega \right)^2}\right)\qquad .\label{zword}
\end{eqnarray}
The choice
\begin{equation}
\bar{{\cal H}}_2= -h_\bot^2 a \frac{\left( 1 - S_z^2 \right) \left( \Omega^2
 + \omega^2 \right) } { 8 \left( \Omega^2 - \omega^2  \right)^2}+
 h_\bot^2 \frac{S_z \Omega }{ 4\left( \Omega^2 -\omega^2 \right)}
\end{equation}
ensures that the solution possesses no secular terms.
The generator becomes
\begin{eqnarray}\label{myt}
    w_2&=& -h_\bot^2 \frac{ \Omega\:S_z \sin\left(2 \omega t\right)}
         {4 \omega \left( \Omega^2 -\omega^2 \right) }+ h_\bot^2
         a \frac{
         \left( 1- S_z^2\right)\left(\Omega^2+ \omega^2
         \right) \sin\left(2 \varphi\right)}{8\Omega\left( \Omega^2 -
         \omega^2 \right)^2}\nonumber\\
      &&+h_\bot^2 a \frac{\left(1-S_z^2\right)\left(\Omega^2+ \omega^2
         \right) \sin\left(2 \omega t\right)}{8\omega\left(
         \Omega^2 - \omega^2 \right)^2}\nonumber\\
      &&+h_\bot^2 a \left(1-S_z^2\right)\left(\frac{\sin\left(2\varphi - 2
         \omega t\right)}{16\left(\Omega-\omega\right)^3} + \frac{
         \sin \left( 2 \varphi + 2\omega t \right) } {16 \left( \Omega
         +\omega \right)^3} \right)\quad.
\end{eqnarray}
Clearly an additional resonant small denominator appears at
$\Omega(S_z)=0$. The full transformation reads up to the second order
in view of
eq.(\ref{myia})
 \begin{equation}\label{myh}
{\bf T} f =1-\varepsilon \left[ w_1, f\right]-
\frac{\varepsilon^2}{2} \left( \left[ w_2, f \right]-
\left[ w_1, \left[ w_1, f \right]\right] \right) \quad .
\end{equation}
and reflects the first and second order resonances.

In analogy to the procedure described in paragraph 2.1.2 the new resonant
denominator can be removed by a different choice for the new
Hamiltonian. The appropriate expression is given by
the time average of the right hand side of
eq.(\ref{myg})
\begin{eqnarray}\label{myf}
\bar{\cal{H}}_2&=&h_\bot^2 a \frac{ \left( \cos \left( 2 \varphi\right)- 1
        \right) \left( 1 - S_z^2 \right) \left(\Omega^2
	 + \omega^2 \right) } { 8 \left( \Omega^2 -
         \omega^2  \right)^2} \nonumber\\
      &&+h_\bot^2 \frac{S_z\;\Omega } { 4\left( \Omega^2
         -\omega^2 \right)} \quad .
\end{eqnarray}
The corresponding solution for the generator remains regular near the
second order resonance
\begin{eqnarray}
   w_2&=& -h_\bot^2 \frac{ \Omega S_z \sin\left(2 \omega t
         \right) }   { 4 \omega \left( \Omega^2 -
         \omega^2\right)}+h_\bot^2 a
         \frac{ \left( 1 - S_z^2 \right) \left(
         \Omega^2 +  \omega^2 \right) \sin \left( 2 \omega t
         \right)
         }{8 \omega \left(\Omega^2 - \omega^2
         \right)^2 }\nonumber\\
      &&+h_\bot^2 a \left(1-S_z^2\right)\left(\frac{\sin\left(2\varphi - 2
         \omega t \right) } { 16 \left( \Omega- \omega
	 \right)^3}  + \frac{ \sin \left( 2 \varphi + 2\omega t \right) }
	 { 16 \left(\Omega + \omega \right)^3 } \right)
 \quad .
\end{eqnarray}
It should be mentioned that one has to pay for the removal of the
resonance at $\Omega=0$ by the explicit occurrence of the angle variable
$\varphi$ in the expression (\ref{myf}) for $\bar{{\cal H}}_2$.

Considering now the transformed Hamiltonian and recalling that
$\bar{{\cal H}}_1$ vanishes (cf. paragraph 2.1.1) we have in view
of eq.(\ref{myf})
\begin{equation}
\bar{\cal H}={\cal H}_0+\varepsilon^2 \;\bar{\cal H}_2 \quad .
\end{equation}
Expanding this Hamiltonian
near the second order resonance $\Omega(S_z)=0$
in powers of small deviations $\Delta S_z$ from
the resonant surface we obtain again
the Hamiltonian of the pendulum
in the lowest non vanishing order
\begin{equation}
  \bar{{\cal H}}=-\frac{a}{2} \left(\Delta S_z \right)^2 - \varepsilon^2 \;
  h_\bot^2 \frac{H_z^2- a^2 }{8 a \omega^2}  \cos 2 \varphi
  \qquad.
\end{equation}
In contrast to the previous case of the first order resonance the width of
the resonant region scales with the amplitude of the driving field
$\varepsilon h_\bot$.
\subsubsection{Global Analysis \label{global}}
The discussion of the previous paragraphs has shown that one can detect
resonances in the spin system and can describe the
phase space structure near the resonances approximately. In addition
it would be of course desirable to
gain a global overview over the phase space structure. This goal
however cannot
be achieved with the help of the local analysis of the previous
sections because the generators contain singularities away from the
resonances (cf. eq.(\ref{myx})).

In order to get an overview of the
phase portrait of the Hamiltonian system
we follow a method which has been proposed by Dunnet at al. \cite{Dune}.
Its main idea is quite simple and is based on the construction of an
approximate constant of motion.
Suppose that the original Hamiltonian has been transformed in some
order to a system whose Hamiltonian depends only on the action variable
$\bar{S_z}$. Then an arbitrary function of the action $\bar{f}(\bar{S_z})$
yields a constant of motion of the transformed system. Hence the
counter image ${\bf T}^{-1} \bar{f}$ is also a constant of motion and
contains the topology of the phase space curves in the original
variables. This naive view is in general meaningless because of the
small denominators involved.
But the construction can be performed with the
singular expression in every order of the perturbation theory if the
constant of motion $\bar{f}$ is suitably chosen.

To be definite consider the perturbation expansion up to second order
(cf. eq.(\ref{myh})). Formal application of expansion (\ref{myib})
leads to
\begin{equation}\label{myk}
f=T^{-1}\bar{f}=\bar{f}+\varepsilon \left[ w_1,\bar{f}\right]+
\frac{\varepsilon^2}{2} \left( \left[ w_2,\bar{f}\right]+
\left[ w_1, \left[ w_1,\bar{f} \right]\right] \right) \quad .
\end{equation}
This expression contains singularities at the
resonances $\Omega(S_z)=0,\pm\omega$. If however
the function
$\bar{f}$ is chosen in such a way that the zeros of its derivative
cancel all these singularities the
expression remains regular in the whole phase space. The authors of
ref.\cite{Dune} suggested this expression to be
a reasonable approximation for
a global invariant of the system.

In principle one can use any function $\bar{f}$ which
satisfies this constraint. For graphical purposes it is however
convenient to
use an expression whose
numerical values are of the same order of magnitude in a region of phase
space as large as possible. We therefore
start from the following expression
\begin{equation}\label{myj}
\bar{f}(\bar{S}_z)= \int_0^{\bar{S}_z}\cos^3
\left( \frac{\pi\Omega(x) }{2\omega} \right)
\Omega(x)\,dx\quad.
\end{equation}
Combining eqs.(\ref{mys}), (\ref{myt}), and (\ref{myk}) the
invariant up to second order perturbation theory is obtained
after some tedious calculation. We refrain from writing down the
lengthy result explicitly.

The level lines of the invariant constructed in this manner provide
a good approximation
of the phase space structure. We give plots of these level lines at
fixed time $t$ which is equivalent to a Poincare section at the same
time. Fig.2 \marginpar{$<$ Fig.2} shows the lines for the parameter values
$\varepsilon h_\bot=0.01$, $a=1$, $H_z=0.5$ and $\omega=0.2$. Resonances
of first and second order are clearly visible. The size of these
resonances is in accordance with the discussion given in the preceding
paragraphs. Additionally second order islands can be recognized near
the first order resonance. But these structures are reproduced only
qualitatively because they depend on the expansion parameter in a
highly nonanalytical way (cf. paragraph 2.1.3 and ref.\cite{McNa}).
If the amplitude of the driving field is increased the level lines which
cross the phase space are destroyed. Fig.3 \marginpar{$<$ Fig.3}
shows an example for the parameter values
$h_\bot=0.025$, $a=1$, $H_z=0.5$ and $\omega=0.2$. This effect
may be attributed to an overlap of low order resonances and
signals the destruction of KAM tori. However higher order perturbation
calculations would be necessary to obtain more reliable results.
\subsection{Numerical Results \label{num}}
We confirm the analytical perturbational calculations of the previous
section by direct numerical simulations of the full Hamiltonian
dynamics. For that purpose the system has been integrated
by using the
subroutine D02BAF of the NAG library. Poincare plots have been
computed by integrating over 100 periods of the driving field starting
from 400 initial conditions distributed
uniformly at $\varphi=0,\pm\pi/2,\pi$.
Fig.4 \marginpar{$<$ Fig.4} shows the Poincare plot
for the parameter values used in Fig.2. Good agreement is observed. But
e.g. third order resonances are found in the numerical simulation
being not incorporated in the perturbative approach which is
of second order only.
Fig.5\marginpar{$<$ Fig.5} contains data at the parameter values used
in Fig.3. Obviously
KAM tori are destroyed by resonance overlap.
The agreement between the perturbation expansion (cf. Fig.3) and the
numerical simulation is merely qualitative. It signals that the
validity of the expansion breaks down if the system is
chaotic in large regions of the phase space.
\section{The Circularly Polarized Driven Dissipative System\label{gamrot}}
We will now study the dynamics of the dissipative system which is
fundamentally different from the case analysed in the previous section.
It is our intention to give a rather complete survey over the
dynamical behaviour in a simple but highly nontrivial situation.
We consider
a system being driven by a rotating transversal field.
The effective field reads
\begin{equation}\label{4.1}
{\bf H}_{eff}({\bf S},t)=
\left(H_z-a S_z \right) {\bf e}_z + h_\bot \left(\cos
\left(\omega t \right) {\bf e}_x+\sin \left(\omega t
\right){\bf e}_y \right)\qquad.
\end{equation}
The explicit time dependence of the equations of motion (\ref{diss}) may be
eliminated using a transformation to a rotating frame because of the
chiral symmetry of the dynamics. We obtain the autonomous equations
of motion
\begin{equation}\label{gamrotbe}
\frac{d {\bf S}}{dt}=- {\bf S}\times \left({\bf H}_{eff} + \Gamma
\;{\bf S}\times \left({\bf H}_{eff}+\omega {\bf e}_z\right)\right)
\end{equation}
where
\begin{equation}
{\bf H}_{eff}= \left(H_z-a S_z- \omega \right){\bf e}_z+h_\bot
{\bf e}_x
\end{equation}
denotes the time independent effective field in the rotating frame.
It is
the physical nature of the driving field which is responsible for
the different magnetic fields in the reversible and the Gilbert term.
This structure prevents eq.(\ref{gamrotbe}) from possessing a
Lyapunov function and ultimately causes the complex behaviour of its
solutions.
On the other hand the modulus of the magnetization is preserved
in the course of time
so that the phase space is the two dimensional sphere.
This fact enables us to keep the discussion to a great extent
analytical but at the same time prevents the system from becoming chaotic.

As a constant factor can be absorbed into the time scale we restrict
the subsequent treatment to the case $|a|=1$ without loss of generality.
We will start our analysis with a brief discussion of the fixed points.
Subsequently local and global bifurcations will be analysed which will
result in a complete overview of the dynamics of the system.

\subsection{The Fixed Points}
Even the calculation of the fixed points gives rise to
a system of algebraic equations in general which cannot be solved without
resorting to numerical methods. But our fixed point problem
(cf. eq.(\ref{gamrotbe}))
can be reduced to a single  algebraic equation of fourth order
\begin{equation}\label{mym}
   \left(1-S_z^2 \right)\left(\left(a S_z -\Delta\right)^2+\gamma^2 h_\bot^2
      S_z^2\right)-h_\bot^2 S_z^2=0 \quad .
\end{equation}
Here the new parameters
\begin{eqnarray}
\Delta &:=& H_z-\frac{\omega}{1+\Gamma^2}\\
\gamma &:=&\frac{ \omega } {h_\bot} \frac{ \Gamma }{ 1+ \Gamma^2}
\end{eqnarray}
have been introduced and will be used in the sequel. The remaining
components are determined via
\begin{eqnarray}
S_x &=& \frac{\Delta- a S_z}{S_z h_\bot} \left(1-S_z^2 \right)\\
S_y &=&-\gamma \left(1-S_z^2\right) \quad .
\end{eqnarray}
It is worthwhile to mention that because of eq.(\ref{mym}) the system
has at least two and at most four fixed points. The explicit discussion is
postponed to the next sections.

\subsection{Method of Analysis}
Before we enter the discussion of the various bifurcation diagrams
it is useful
to give a brief survey over the symmetries of the system under
consideration. It is easy to check that the equations of motion
are invariant with respect to the following two transformations
\begin{eqnarray}
{\bf T}_1:&&\Delta\rightarrow -\Delta \quad
\gamma\rightarrow - \gamma  \quad a\rightarrow -a\quad  h_\bot
\rightarrow -h_\bot\quad  t\rightarrow -t\\
{\bf T}_2:&& \Delta \rightarrow -\Delta \quad
\gamma\rightarrow -\gamma \quad S_z \rightarrow -S_z \quad
S_y=-S_y \quad .
\end{eqnarray}
Taking these symmetries into account we can restrict the following
bifurcation analysis to the case $a=-1$ and $\gamma >0$.
In addition we will not analyse the full four dimensional
parameter space ($\Gamma$, $\Delta$, $\gamma$, $h_\bot$). We restrict
ourselves to the fixed value of the damping constant $\Gamma=0.1$.

Let us first describe the procedure which is applied to treat
the local codimension one bifurcations
(Hopf and saddle node bifurcations). It is convenient to perform this
analysis by referring explicitly to the fact that the phase space is
two dimensional. We rewrite eq.(\ref{gamrotbe})
in terms of the complex variable
$Z= \left(S_x+ i S_y \right)/ \left(1+S_z \right)$. This corresponds to
the well known stereographic projection of the sphere to the plane
(cf. ref.\cite{Bala})
\begin{equation}\label{myn}
\frac{d Z}{dt} = f(Z) := \left( i -\Gamma \right)
\left( i \gamma h_{\bot} Z + \Delta Z -
a Z \frac{1-|Z|^2}{1+|Z|^2} - \frac{h_{\bot}}{2} \left( 1- |Z|^2\right)
\right)
\quad .
\end{equation}
Saddle node bifurcations of the fixed points are determined by a single
vanishing eigenvalue of the linearized system i.e. by the
equations\footnote{Inspection of the polynomial eq.(\ref{mym})
shows that this condition coincides with a degeneracy of its zeroes.}
\begin{equation}\label{myo}
f(Z)=0, \qquad \det(Df(Z))=0 \quad .
\end{equation}
The solutions of this set of algebraic equations define codimension one
manifolds in parameter space on which saddle node bifurcations take
place. With the help of the program PITCON \cite{pit} these manifolds
can be computed easily by a continuation technique. In the same way Hopf
bifurcations are detected. Owing to the dimensionality of the phase space
they are determined by the equations
\begin{equation}\label{myoa}
f(Z)=0, \qquad \mbox{Tr}(Df(Z))=0 \quad .
\end{equation}
In addition using a well known formula for the coefficient
of the third order in the Hopf normal form \cite{GuHo} we decide whether the
bifurcation is sub-- or supercritical. We stress that
our explicit knowledge of the number of fixed points enables us to
detect all local codimension one bifurcations.

The discussion of the local codimension one bifurcations will show
that our system has at most one saddle point.
Homoclinic bifurcations occurring eventually in connection with this
point
have been investigated using a numerical computation
of the distance between its stable and unstable manifolds.
This method seems to be more suitable than the quest for limit cycles
as suggested by Doedel (ref.\cite{doed}).
The respective codimension one bifurcation sets are again computed via PITCON.

Finally we have concentrated on the investigation of saddle node
bifurcations of limit cycles. It is well known that the proper
bifurcation manifold is born in a degenerated Hopf bifurcation.
Using the program AUTO \cite{doed} we have been able to compute
the saddle node bifurcation manifold of limit cycles also.
\subsection{Bifurcation Diagrams}
The results of our investigations are summarized in Figs.6,13--16.
In order to get insight into the structure of the full three dimensional
parameter
space we show cross sections
for several values of $h_\bot$. Let us first
describe in detail the situation for
$h_\bot=0.1$ (cf. Fig.6) \marginpar{$<$ Fig.6}. Apart from the
codimension one manifolds we find the following bifurcations\footnote{For a
detailed explanation of these basic bifurcations the reader is
referred to \cite{GuHo}} of codimension two:
\begin{itemize}
\item two cusp points at $C_1$ and $C_2$
\item one degenerated Hopf bifurcation at $H$ where a saddle node
bifurcation for limit cycles meets a Hopf bifurcation
\item two Arnold--Takens--Bogdanov  bifurcations at $A_1$ and $A_2$
where saddle node, Hopf, and homoclinic bifurcations meet
\item two saddle node connections at $S_1$ and $S_2$
\item one degenerated homoclinic
bifurcation\footnote{For a description of this bifurcation
see e.g.~\cite{jo}.}  at $B$ where a saddle node bifurcation line
of limit cycles meets
a homoclinic bifurcation. The lines to the left of $B$
cannot be resolved on the scale
of Fig. 6 (cf. the small insert).
\end{itemize}
The local and global codimension one lines divide the parameter
space in ten regions. The typical dynamical behaviour in each region
and on the boundaries is described in the sequel:
\begin{itemize}
\item In region I there exist only one stable and one
unstable fixed point.
\item[--] The saddle node bifurcation line,
which separates region I from
region II,  generates a saddle point and a stable fixed point.
\item Region II contains four fixed points, one saddle point, one
unstable fixed point and two stable fixed points.
\item[--] The saddle node bifurcation line,
which separates
region II from region III, destroys a stable fixed point and the saddle
point.
\item In region III there exist one stable and one unstable fixed point.
\item[--] The subcritical Hopf bifurcation line,
which separates region III from
region IV, generates an unstable limit cycle. The unstable fixed point
becomes stable.
\item region IV contains two stable fixed points and one unstable limit
cycle.
\item[--] The subcritical Hopf bifurcation line,
which separates region IV from
region I destroys the unstable limit cycle. One stable fixed point becomes
unstable.
\item[--] The supercritical Hopf bifurcation line,
which separates region IV from
region V, generates a stable limit cycle and turns one stable fixed point
into an unstable fixed point.
\item In region V there exist a stable and an unstable fixed point
and a stable and an unstable limit cycle.
\item[--] The limit cycles just mentioned  are destroyed
at the
saddle node bifurcation line for
limit cycles, which separates region V from region I.
\item[--] The saddle node bifurcation line, which separates region V from
region VI (cf. the small insert in Fig.6),
generates a stable fixed point and a saddle point. Moreover this bifurcation
destroys the stable limit cycle.
\item Region VI contains four fixed points (one saddle, one unstable and
two stable fixed points) and an unstable limit cycle.
\item[--] The homoclinic bifurcation line, which separates region VI from
region II,
destroys the unstable limit cycle.
\item Region II contains the same fixed points as region VI but no
limit cycle.
\item[--] The saddle node bifurcation line, which separates
region VI from
region IV, destroys an unstable fixed point and the saddle point.
\item[--] The subcritical Hopf bifurcation line, which separates region VI
from region VII, destroys the unstable limit cycle and turns a stable
fixed point unstable.
\item Region VII contains only four fixed points (one saddle, two unstable
and one stable fixed point).
\item[--] The saddle node bifurcation line, which separates region VII from
region III, destroys an unstable fixed point and the saddle point.
\item[--] The saddle node bifurcation line, which separates region V from
region VIII, generates a stable fixed point and a saddle point.
\item Region VIII contains four fixed points (one saddle point, one
unstable and two stable fixed points) and two limit cycles (one stable
and one unstable)
\item[--] The homoclinic bifurcation line, which separates region VIII from
region VI, destroys the stable limit cycle.
\item[--] The supercritical Hopf bifurcation line, which
separates region VIII from region IX, destroys the stable limit cycle
and turns an unstable fixed point stable.
\item Region IX contains four fixed points (one saddle and three stable
fixed points) and one unstable limit cycle.
\item[--] The saddle node bifurcation line, which separates region IX from
region IV, destroys a stable fixed point and the saddle point.
\item[--] The homoclinic bifurcation line, which separates region X from
region VI,
generates a stable limit cycle.
\item Region X contains the same fixed points as region VI but two
limit cycles (one stable and one unstable).
\item[--] The saddle node bifurcation line for limit cycles, which separates
region X
from region II destroys both limit cycles contained in region X.
\end{itemize}
Phase portraits for typical parameter values in the regions II and V--IX
are displayed in Fig.7\marginpar{$<$Fig.7}.

Our bifurcation diagram has an intricate structure especially as far
as the two homoclinic bifurcation lines
$\overline{A_1 S_1}$ and $\overline{A_2 S_2}$  are concerned. The insert
of Fig.6 which resolves these curves on a finer scale shows that they end up
in different points $S_1$ and $S_2$. This is a consequence of the fact that
the two bifurcations in question play essentially different rolls with respect
to the global phase portrait of the system (cf. Fig.8).
\marginpar{$<$ Fig.8}
Fig.8a shows a phase portrait on the line $\overline{A_1 S_1}$. One
realizes that the homoclinic orbit encloses one unstable fixed point.
Fig.8b shows the same phase portrait on the other homoclinic bifurcation
line $\overline{A_2 S_2}$. In this case the homoclinic orbit encircles two
fixed points. Both situations cannot be deformed smoothly into each other
and are therefore topological distinct\footnote{Of course one has to be
careful when speaking about the interior of a curve because our phase
space is a sphere and not the plane. In the present case we define the
interior of the homoclinic orbit as those part of the phase space which
contains the angle between the stable and the unstable eigendirection
of the saddle point.}. We add that in Figs.8a and 8b the stable manifold
of the saddle point may be thought of as employing just its two
different branches of the unstable manifold in building the two
homoclinic orbits.

In addition our bifurcation diagram reflects the topology of the phase
space. For example region IV is bounded by two Hopf bifurcation lines.
The upper line generates an unstable limit cycle. It is destroyed
if one crosses the lower subcritical Hopf bifurcation line. As both
Hopf lines are connected to different Arnold--Takens--Bogdanov points
the fixed points involved in the Hopf bifurcations are different.
Hence the limit cycle wanders continuously from one fixed point to the other.
Such a scenario is impossible if the phase space is a plane, e.g.
for a mechanical oscillator.

If we change the third bifurcation parameter $h_\bot$ the bifurcation
diagram
changes smoothly. However some qualitative modifications occur.
Using the same notations
as in the previous analysis we describe the major
differences. Increasing
$h_\bot$ the Hopf bifurcation line which begins at $A_2$ flattens
and finally divides region VIII and IX into two parts
(cf. Fig.9). \marginpar{$<$Fig.9}
In contrast to regions VIIIa and IXa
regions VIIIb and IXb contain an additional
unstable limit cycle. At slightly higher
values of the driving field $h_\bot$ the same phenomenon occurs in
region V (cf. Fig.10). \marginpar{$<$Fig.10}

A dramatic change in the bifurcation diagram occurs
at the Arnold--Takens--Bogdanov point $A_2$
by increasing the driving field further.
Fig.11
\marginpar{$<$Fig.11} shows the diagram for $h_\bot=0.8$.
The major differences to the previous case (cf. Fig.10) are
\begin{itemize}
\item The Hopf bifurcation line starting at the point $A_2$ has become
supercritical.
\item The homoclinic bifurcation line lies above the Hopf bifurcation line.
\item The saddle node bifurcation for limit cycles now ends up in
a degenerated Hopf bifurcation point instead of a degenerated homoclinic
bifurcation point.
\end{itemize}
Hence the Arnold--Takens--Bogdanov bifurcation has changed its type.
In terms of the normal form
\begin{eqnarray}
\dot{x}&=&y\label{mypa}\\
\dot{y}&=&a \:x^2+b \:x \:y \label{mypb}
\end{eqnarray}
this bifurcation can be attributed to a change in sign of the coefficient
$b$. This degenerated Arnold--Takens--Bogdanov bifurcation
is discussed in the mathematical literature on an abstract level
\cite{Du}.
In order to keep the discussion self contained as well as accessible to
physicists we have developed a much more elementary approach
which nevertheless yields the complete bifurcation diagram of this
codimension three bifurcation point. It is given in the appendix.

An additional bifurcation of (at least) codimension three occurs at
$h_\bot=1$. If we approach this point the saddle node bifurcation lines
collide at $\gamma=0, \Delta=0$ and generate
two new cusp points $C_3$ which initially coincide (cf. Fig.12).
\marginpar{$<$Fig.12} The center manifold of this bifurcation is
one dimensional because only a single  eigenvalue vanishes.
On a further increase of the driving field the cusp points $C_3$
separate. Fig.13 \marginpar{$<$Fig.13} shows the bifurcation diagram.

\section{Conclusion}
Our analysis of a driven spin system has covered both aspects of the dynamics,
the Hamiltonian as well as the dissipative case. The undamped linearly
polarized driven system discussed in section 2 has shown the typical behaviour
of a nonintegrable Hamiltonian system. Our analysis has
revealed that only
a finite number of resonances occurs in each order of the
perturbation expansion. This is a consequence of the direct coupling
of the driving field to the action angle variables (cf. eq. (\ref{4.5}))
which are in a certain sense the natural variables of the dynamical
system. In this respect our system differs from e.g. mechanical
oscillators where higher harmonics come into play even at low
order resonances. The local phase space structure near these resonances
including their width was described by a pendulum equation.
In addition the phase portrait has been analysed using a
global perturbative approach. It is in good agreement with numerical
simulations.

The investigation of the dissipative system subjected to a circularly
polarized driving field has revealed a surprisingly rich bifurcation
scenario. Beside local (saddle node and Hopf) and global bifurcations
(homoclinic and saddle node bifurcations of limit cycles) of codimension one
we have found five different bifurcations of codimension two and even two
bifurcations of codimension three. From our analysis, which has been
performed to a large extent analytically, we draw the conclusion that
our bifurcation diagram is complete. We notice that for a closely
related model, i.e. an antiferromagnet made up of two homogeneous
sublattices, a partial bifurcation diagram has been established in
\cite{benno}.

It is widely believed and supported by our results that the
Hamiltonian and the dissipative dynamics are entirely different. But
the limit of weak damping and the emergence of dissipative dynamics from
the Hamiltonian phase space structure is poorly understood. The system
analysed in this article provides a physical model to study this issue.

Our investigations constitute a necessary prerequisite for a systematic
study of spatially inhomogeneous states of the full Landau--Lifshitz--Gilbert
equation. A rather complex behaviour even under the very
restrictive assumption of homogeneous magnetization was found.
We are convinced that the full spatially inhomogeneous equation,
including exchange as well as dipolar interaction, provides the correct
theoretical description of strongly driven
magnetic systems far from equilibrium. Work in this direction is in
progress and will be published elsewhere.

\section*{Acknowledgements}
One of the authors (W.J.) is indebted to the ''Deutsche
Forschungsgemeinschaft'' for support by a ''Habilitandenstipendium''.
This work was performed within a program of the Sonderforschungsbereich
185 Darmstadt--Frankfurt, Germany.

\section*{Appendix: The degenerated Arnold--Takens--Bog\-da\-nov Bifurcation
\label{ATB}}
We present here an elementary discussion of the degenerated
Arnold--Takens--Bog\-da\-nov bifurcation.
The mathematical background for this codimension three bifurcation was
worked out in ref.\cite{Du}.
Hence we skip the formal proof of the universality of the unfolding but
concentrate on the computation of the bifurcation manifolds in the full three
dimensional parameter space.

We consider the case that in the well known normal form of the
Arnold--Takens--Bogdanov bifurcation (\ref{mypa},\ref{mypb})
a degeneracy of the second
order terms occurs, which means that the coefficient $b$ vanishes. In order
to establish the normal form of this higher order codimension point we
start from the general equations of motion
\begin{eqnarray}
\dot{x}_1&=\quad f_1({\bf x})&= \quad x_2 + \sum_{j=3}^{\infty} \sum_{i=0}^{j}
a_{ij}x_1^ix_2^{(j-i)}\label{1}\\
\dot{x}_2&=\quad f_2({\bf x})&= \quad x_1^2 +\sum_{j=3}^{\infty} \sum_{i=0}^{j}
b_{ij}x_1^ix_2^{(j-i)}\label{1a}
\end{eqnarray}
which coincide up to  second order with the degenerated situation just
described.  They contain all possible contributions beyond the second order.
In the spirit of normal form calculations we are looking for
a coordinate transformation
${\bf x}={\bf h}({\bf y})$, ${\bf h}({\bf 0})={\bf 0}$
including a rescaling of the
time $t\to \zeta t$ such that the system (\ref{1},\ref{1a})
becomes ``as simple as
possible'' \cite{GuHo}. In the new coordinates we get
\begin{equation}\label{dy}
\dot{{\bf y}}=  {\bf g} \left({\bf y} \right)=\zeta\left(D{\bf h}
\left({\bf y}
\right) \right)^{-1} {\bf f} \left({\bf h} \left({\bf y} \right) \right)\quad .
\end{equation}
Substituting the expansions
\begin{eqnarray}\label{yp}
h_1 &=& c_{11}y_1 + \sum_{j=2}^{4} \sum_{i=0}^{j} c_{ij} y_1^i
y_2^{j-i}\nonumber\\
h_2 &= & d_{11}y_2 + \sum_{j=2}^{4} \sum_{i=0}^{j} d_{ij} y_1^i y_2^{j-i}
\end{eqnarray}
into equation (\ref{dy}) and keeping terms up to fourth order
we determine the coefficients $c_{ij}$ and $d_{ij}$ in such a way that
as many terms as possible vanish in equation (\ref{dy}). The necessary
algebraic manipulations are quite extensive. They were performed with the
help of a computer program applying symbolic mathematics \cite{Wo}.
We end up with the normal form
\begin{eqnarray}\label{normal}
\dot{y}_1 & = & y_2 + {\cal O} \left(  \left|{\bf y}
\right|^5\right)\nonumber\\
\dot{y}_2 & = &  y_1^2 + \alpha\; \;y_1^3 + y_1^3 \;y_2 +{\cal O} \left(
\left|{\bf y} \right|^5\right)
\end{eqnarray}
where $\alpha$ is a constant of order unity which remains undetermined.

In extension and analogy to the two parameter unfoldings of the standard
Arnold--Takens--Bogdanov bifurcation \cite{GuHo} we now introduce the
following three parameter unfolding
\begin{eqnarray}
\dot{y}_1 & = & y_2 + {\cal O} \left(  \left|\bf{y} \right|^5\right)
\label{unfolding}\\
\dot{y}_2 & = & \mu_1+\mu_2 \;y_2+\mu_3 \;y_1\;y_2+ y_1^2 + \alpha  \; y_1^3 +
  y_1^3\; y_2 +{\cal O} \left(  \left|\bf{y} \right|^5\right) \quad.
\label{ufa}
\end{eqnarray}
The bifurcation manifolds in the three
dimensional parameter space $(\mu_1,\mu_2,\mu_3)$
will be computed in the sequel.

We begin our discussion with an analysis of the local codimension one
bifurcations. Following the lines of section 3.2 we obtain from
eqs.(\ref{myo}) easily the saddle node bifurcation manifold
\begin{equation}\label{myq}
\mu_1^{sn}=0 \quad .
\end{equation}
Evaluating eqs.(\ref{myoa}) a little algebra yields the Hopf bifurcation
manifold
\begin{equation}\label{hopf}
\mu_2^{\mbox{\tiny Hopf}}(\mu_1,\mu_3)=\sqrt{-\mu_1} \left(\mu_3 -\mu_1
\right)+{\cal O} \left(\mu_3\;\mu_1,\mu_1^2
\right)\quad.
\end{equation}
Fig.14 \marginpar{$<$ Fig.14} shows both manifolds in the three dimensional
parameter space.

To study the global bifurcations a rescaling is performed
which differs from the blowing up used in the treatment of the
ordinary Arnold--Takens--Bogdanov bifurcation \cite{GuHo}. One has to
choose
\begin{equation}\label{trafo}
y_1=\varepsilon^2 \;u,\quad y_2=\varepsilon^3 \; v,\quad \tau=\varepsilon \;
t,
\end{equation}
\[ \mu_1=\varepsilon^4 \;\nu_1,\quad
\mu_2= \varepsilon^6 \; \nu_2, \quad \; \mu_3=\varepsilon^4 \;
\nu_3\quad.
\]
in order to ensure that the rescaled system becomes Hamiltonian in
lowest order and that all non Hamiltonian contributions scale with the
same order of magnitude of $\varepsilon$.
We obtain from eqs.(\ref{unfolding},\ref{ufa})
\begin{eqnarray}
\dot{u}&= &\;v + {\cal O} \left(\varepsilon^6 \right)\label{un}\\
\dot{v}&= &\nu_1 +u^2 + \varepsilon^{2}\; \alpha \; u^3
+ \varepsilon^{5} \;\nu_2\;v+ \varepsilon^{5} \;\nu_3\;u \; v
+\;\varepsilon^{5}\;u^3\;v+{\cal O}
\left(\varepsilon^6 \right)\quad . \label{una}
\end{eqnarray}
where the dot now denotes the derivative with respect to the new time
$\tau$. We note that the term proportional to $\varepsilon^2$ is Hamiltonian.
We are now dealing with a {\it four} parameter problem where the parameters
$\nu_1$, $\nu_2$, $\nu_3$ are of order
${\cal O}(1)$ and $\varepsilon$ is small.
In the limit $\varepsilon \to 0$ eqs.(\ref{un},\ref{una}) yield an integrable
Hamiltonian system with the Hamiltonian
\begin{equation}\label{ham}
H(u,v)=\frac{v^2}{2}-\nu_1u- \frac{u^3}{3}\quad .
\end{equation}
Fig.15 \marginpar{$<$ Fig.15} shows the corresponding phase portrait.
It exhibits closed orbits and a homoclinic loop $\Gamma_0$ with the
energy $H(u,v)=\frac{2}{3}\sqrt{-\nu_1^3}$.

The singular transformation (\ref{trafo}) has blown up the degenerated fixed
point into a Hamiltonian system because the limit $\varepsilon \to 0$
implies $\mu_1,\mu_2,\mu_3 \to 0$. By perturbing the
solutions of the Hamiltonian system (\ref{ham}) we are able to uncover
the behaviour of eqs.(\ref{un},\ref{una}) for parameter values
$\mu_1,\mu_2,\mu_3$ close to the origin.

Homoclinic bifurcations occur at those parameter values
$\nu_1,\nu_2,\nu_3$ for which the homoclinic orbit of the
Hamiltonian system persists if the perturbation is switched on.
This problem can be treated analytically by Melnikovs method
(e.g. ref.\cite{GuHo}).
The (nondegenerated) zeroes of the time independent Melnikov function
\begin{eqnarray}
M(\nu_1,\nu_2,\nu_3)&=&\int_{\Gamma_0(\nu_1)}  \left(\nu_2\:v+\nu_3 \:u\:v+
u^3\:v \right)\;du\nonumber\\
&=& \frac{8  \sqrt[4]{-4\nu_1^5}}{385} \left(-309 \sqrt{-\nu_1^3} + 231\:
\nu_2
- 165\: \nu_3\:\sqrt{-\nu_1}  \right)
\end{eqnarray}
signal the occurrence of the homoclinic bifurcation. In view of the scaling
(\ref{trafo}) this condition reads
\begin{equation}\label{17}
\mu_2^{hc}=\frac{103 \sqrt{-\mu_1^3}+ 55 \:\mu_3\:\sqrt{-\mu_1}}{77}\quad.
\end{equation}
The surfaces at which the   Hopf, saddle node  and homoclinic
bifurcations take place are depicted in Fig.16\marginpar{$<$ Fig.16}.

Finally we show that a further global bifurcation, a saddle node bifurcation
of limit cycles, comes into existence for the degenerated
Arnold--Takens--Bogdanov bifurcation. The subharmonic Melnikov function
\begin{equation}\label{smel}
M_s(\nu_1,\nu_2,\nu_3,E)=\int_{\Gamma(E,\nu_1)}  \left(\nu_2\:v+\nu_3 \:u\:v+
u^3\:v \right)\;du
\end{equation}
provides an appropriate tool for studying this issue \cite{GuHo}. Here
$\Gamma(E,\nu_1)$ denotes a trajectory of the Hamiltonian system (\ref{ham})
with energy $E$. If this function has a nondegenerated zero
for a certain value of the energy then the corresponding periodic orbit
persists in the perturbed system (\ref{un},\ref{una}). Moreover a doubly
degenerated zero of $M_s$ indicates
a saddle node bifurcation of limit cycles at the corresponding parameter
value. Hence the saddle node bifurcation manifold is determined by
the equations
\begin{eqnarray}
0 = M_s(\nu_1,\nu_2,\nu_3,E) &=& \sqrt{2}\int_a^b \frac{ \left(\nu_2\:u+
\nu_3 u^2/2+u^4/4 \right)
\left(u^2+\nu_1 \right)} {\sqrt{E+\nu_1 u+u^3/3}}\;du
\label{suba}\\
0 = \frac{\partial }{\partial E}M_s(\nu_1,\nu_2,\nu_3,E)
&=& \sqrt{2}\int_a^b
\frac{\nu_2+\nu_3 \:u+ u^3}{\sqrt{E+\nu_1 u+u^3/3}}\;du \label{subb}
\quad .
\end{eqnarray}
Here the limits of the integrals are given by the first two zeroes of the
argument of the square root:
\begin{equation}
E+\nu_1 u+\frac{u^3}{3}=\frac{1}{3}(u-a)(b-u)(c-u)\qquad a<b<c\quad .
\end{equation}
We solve the linear system (\ref{suba},\ref{subb}) for the two parameters
$\nu_2$ and $\nu_3$ and obtain a parametric representation of the saddle node
bifurcation surface of the form $ \left(\nu_2,\nu_3 \right) =
{\bf F}(\nu_1,E)$.

Fig.17 \marginpar{$<$ Fig.17} contains all bifurcation manifolds in
a single diagram. We draw attention to the fact that the saddle node
bifurcation manifold for limit cycles meets the Hopf surface in a line
of degenerated (local) Hopf bifurcations, whereas its contact with
the homoclinic surface yields a line of degenerated (global) homoclinic
bifurcations. Both lines represent codimension two bifurcation manifolds.
Joyal \cite{jo} describes  a related phenomenon in a different context.
In addition we recall that the bifurcation scenario described
at the end of section 3.3 is recovered if one considers two dimensional
cross sections (e.g. at constant $\mu_3$) of our three dimensional bifurcation
diagram. In closing we restate that the bifurcation set presented in Fig.17
yields the complete bifurcation diagram of the degenerated
Arnold--Takens--Bogdanov bifurcation.
\newpage
\noindent
\newpage
\noindent
\section*{Figure captions}
\begin{itemize}
\item[Fig.1] Phase portrait of the Hamiltonian at parameter
values $\varepsilon h_\bot = 0.01$, $a=1$, $H_z=0.5$, and $\omega=0.2$.
\item[Fig.2] Level lines of the approximate second order invariant $f$ at
parameter values $\varepsilon h_\bot = 0.01$, $a=1$, $H_z=0.5$,
$\omega=0.2$ and at time $t=0$.
\item[Fig.3] Level lines of the approximate second order invariant $f$ at
parameter values $\varepsilon h_\bot = 0.025$, $a=1$, $H_z=0.5$,
$\omega=0.2$ and at time $t=0$.
\item[Fig.4] Poincar\'e map at time $t=0$ of the full Hamiltonian
system (\ref{4.5}) at parameter values $\varepsilon h_\bot = 0.01$, $a=1$,
$H_z=0.5$ and $\omega=0.2$.
\item[Fig.5] Poincar\'e map at time $t=0$ of the full Hamiltonian
system (\ref{4.5}) at parameter values $\varepsilon h_\bot = 0.01$, $a=1$,
$H_z=0.5$ and $\omega=0.2$.
\item[Fig.6] Bifurcation diagram of eq.(\ref{gamrotbe}) at parameter
values $h_\bot=0.1$ and $\Gamma=0.1$. The bifurcation lines are denoted as
follows: sn: saddle node bifurcation, Ho: supercritical Hopf bifurcation,
ho: subcritical Hopf bifurcation, hc: homoclinic bifurcation, and
sl: saddle node bifurcation for limit cycles. Codimension two points are
labeled by capital letters: $A_i$: Arnold--Takens--Bogdanov bifurcation,
$B$: degenerated homoclinic bifurcation, $C_i$: Cusp bifurcation,
$H$: degenerated Hopf bifurcation, and $S_i$: saddle node connection.
The numbers refer to the regions described in the text.
The region $(\Delta , \gamma) \in [-0.7 , -0.69]\times  [0.39 , 0.395]$
is magnified in the small insert. Note that to the left hand side of the
point $B$ there exists a line of a homoclinic bifurcations ending up
in point $S_2$ and a line of a saddle node bifurcation of limit cycles ending
up in point $H$.
\item[Fig.7] Phase portraits in stereographic projection
for several points in the bifurcation diagram
Fig.6. The numbers refer to the regions in the bifurcation diagram.
The phase portraits show actual solutions of the differential equations.
Parameter values $(\Delta , \gamma)$ are as follows: (II) $(-0.1 , 0.1)$
(V) $(0.75 , 0.5)$, (VI) $(-0.1 , 0.6)$, (VII) $(0.5 , 0.9)$,
(VIII) $(-0.65 , 0.6)$, and (IX) $(-0.65 , 1)$. In all cases there
exists an additional fixed point on the lower hemisphere which is not
displayed.
\item[Fig.8] Phase portraits at the two different homoclinic lines
(a) $\overline{A_1 S_1}$ and (b) $\overline{A_2 S_2}$ (cf. Fig.6). Parameter
values are chosen as (a) $\Delta=-0.6$, $\gamma=0.604$ and
(b) $\Delta=-0.4$, $\gamma=0.4009$. An additional fixed point exists on the
lower hemisphere which is not displayed.
\item[Fig.9] Bifurcation diagram of eq.(\ref{gamrotbe}) at parameter
values $h_\bot=0.2$ and $\Gamma=0.1$. The notation is the same as in Fig.6.
\item[Fig.10] Bifurcation diagram of eq.(\ref{gamrotbe}) at parameter
values $h_\bot=0.5$ and $\Gamma=0.1$. The notation is the same as in Fig.6.
\item[Fig.11] Bifurcation diagram of eq.(\ref{gamrotbe}) at parameter
values $h_\bot=0.8$ and $\Gamma=0.1$. The notation is the same as in Fig.6.
The lower part shows a magnification of the region near the codimension two
point $A_2$.
\item[Fig.12] Bifurcation diagram of eq.(\ref{gamrotbe}) at parameter
values $h_\bot=1$ and $\Gamma=0.1$. The notation is the same as in Fig.6.
The insert shows a magnification of the region $(\Delta,\gamma)\in
[-0.02,0.02]\times [0.2,0.4]$
\item[Fig.13] Bifurcation diagram of eq.(\ref{gamrotbe}) at parameter
values $h_\bot=1.1$ and $\Gamma=0.1$. The notation is the same as in Fig.6.
The insert shows a magnification of the region $(\Delta,\gamma)\in
[-0.0015,0.0015]\times [0.415,0.445]$
\item[Fig.14] Partial bifurcation set of system (\ref{unfolding},\ref{ufa})
containing Hopf and saddle node (sn) bifurcation manifolds.
\item[Fig.15] The phase portrait of the Hamiltonian (\ref{ham}) at
$\nu_1=-1$. $\Gamma_0$ denotes the homoclinic loop.
\item[Fig.16] Partial bifurcation set of system (\ref{unfolding},\ref{ufa})
containing the
Hopf, saddle node (sn) and homoclinic (hc) bifurcation manifolds (cf.
Fig.14).
\item[Fig.17] Complete bifurcation set of system(\ref{unfolding},\ref{ufa})
containing
the Hopf, saddle node (sn) and homoclinic (hc) bifurcation manifolds
as well as the saddle node bifurcation manifold of limit cycles (sl)
(cf. Fig.16).
\end{itemize}
\end{document}